\pdfoutput=1

\documentclass[a4paper,10pt]{article}
\usepackage{amsmath,amssymb}
\usepackage[colorlinks]{hyperref}
\usepackage{eucal}
\scrollmode
\usepackage{graphicx}
\usepackage{epstopdf}
\usepackage{array, longtable}
\usepackage[english]{babel}

\title { Toward a New Model of the Central Engine of GRB\footnote{This article is based on the talk
given on 18.09.2007 at the Fourth Aegean Summer School, 17-22 September 2007, Lesvos, Greece}}
\author {Plamen P. Fiziev\thanks{ E-mail:\,\,\,fiziev@phys.uni-sofia.bg},  Denitsa R. Staicova\thanks{ E-mail:\,\,\,denijane@gmail.com}
\\Department of Theoretical Physics, Sofia University ``St. Kliment Ohridski", \\ 5
James Bourchier Blvd., 1164 Sofia, Bulgaria }
\date {}

\begin{document}
\maketitle

\begin{abstract}
We present new developments of the simple model of the central engine of GRB, proposed
recently. The model is based on minimal assumptions: some rotating compact relativistic object at the center
and stable perturbations of its rotating gravitational field, described by Teukolsky Master Equation.
We show that using nonstandard polynomial solutions to the angular Teukolsky
equation we can describe the formation of collimated jets of various forms. Appearance of imaginary part of the
superradiance-like frequency is established for the first time for pure vacuum black hole jet solutions
of Teukolsky equation.

\end{abstract}

\section{Introduction}

Gamma-ray bursts (GRB) are among the most powerful astrophysical objects in the Universe.
They are considered as highly collimated ($\theta_{jet}\sim 2^\circ-5^\circ$) explosions on cosmic distances
emitting huge amounts of energy in very short periods of time ($\sim$ seconds) \cite{Piran,Meszaros, Zhang, Burrows}.
Despite the increasing amount of  observational data, the theory of GRB
is still far from being clear. One of the major problems is the lack of understanding of
the nature of the so called central engine -- its physical nature and the process involved
in the emission of huge amounts of energy ($\sim 10^{51} - 10^{54}$ erg).

Unexpected feature discovered by the mission SWIFT is the existence of flares
-- a surprising evidence of late time activity of the central engine \cite{Zhang, Burrows, Evans}.
There are clear indications that in some cases the flares are produced by energy
injection by the central engine.
Although number of theories and models have been suggested to explain the light curves of GRB,
the process however remains a mystery that cannot be solved without a good model of the central engine.

The current focus of GRB physics is on the propagation of the emitted matter that can explain
the observed light curves. The most used is the fireball model (for details see \cite{Piran}).
That model, however, cannot explain the observed flares.
Another basic model is the cannonball model of long GRBs \cite{cannonball}.

The central engine -- the physical object producing the GRB -- before SWIFT
epoch usually was considered to be a Kerr \cite{Kerr} black hole (BH)  --
a hypothetical result of the death of a massive star for the long GRB ($T_{90}>2s$)
(see \cite{Zhang}, \cite{Mirabal})
or as part of a binary merger of compact objects (BH-BH, NS-BH, or NS-NS; where NS=neutron star) for short GRB ($T_{90}<2s$).
The first hypotheses seems to be hardly compatible with the observed flares, produced via energy
injection by the central engine.
The second hypotheses was recently refuted by the existing detectors of
gravitational waves \cite{Andromeda}. The latest analysis shows that the short and long GRB may have
a similar central engine (except for its duration) \cite{short_long}.

It was believed that one of the possible models for such engine could be
a Kerr black hole (KBH) in super-radiant mode \cite{ Teukolsky, superradiance} --
the wave analogue of the Penrose process.
The late-time evolution of a perturbation of a Kerr metrics is governed by
the quasi-normal modes (QNM) -- oscillation with complex frequencies that are
determined only by the BH parameters (mass, charge and angular momentum) \cite{QNM, Teukolsky}.
Those frequencies are obtained solving  the Teukolsky Master Equations with certain boundary conditions.
According to the standard theory, supperradiance occurs in the process of scattering of exterior waves on KBH for {\em real} frequencies $\omega<\omega_{critical}=am/2Mr_{_+}$ ($m>0$ -- integer, $r_{_+}>0$ being the event horizon radii) \cite{superradiance}.

Another process which was currently related with the GRB central engine is the Blandford-Znajek process \cite{Blandford-Znajek}
based on electromagnetic extraction of energy from KBH. Sometimes it is considered as an electromagnetic analog to Penrous process.
The realization of this idea has its own achievements and problems (see in  \cite{Blandford-Znajek}).

However, the Penrose process seems to be not efficient enough to
explain the huge amount of promptly radiated energy in relativistic jets ($\sim 10^{41} - 10^{44}$ erg)
\cite{jets}, as shown long time ago by Wald (see in \cite{superradiance}).

In a series of talks \cite{talks}, we have presented a mathematical description of relativistic jets
that can help to penetrate into the mysterious physics behind the central engine. A simple model, based
on novel solutions of the angular Teukolsky equation, however, not regular,
but singular ones \cite{talks, Fiziev0902.1277, RBPF} was proposed.
They seem to describe in natural way the most important feature of the  relativistic jets -- their collimation.
In this paper, we would like to announce the recent results of our numerical simulations.
The main new result is the appearance of a {\em complex}
critical frequency, which play the role of the well known supperradiance ones.
Its imaginary part is of the same order of amplitude as the real part
and yields an exponential abatement of the superradiance-like emission in jets, created by KBH.

\section{A Toy Model of Central Engine}
Our simplified model was already presented in \cite{talks,Fiziev0902.1277}. Thus,
we will omit the details and give only a summary of the basic ideas behind it.
In brief, the jets observed in GRBs require a rotating object -- KBH
or a compact massive {\em matter} object, to produce them.
In the jet's problem we can use the Kerr metric to describe the gravitational
field of that object at least in a very good approximation \cite{Fiziev0902.1277},
since the visible jets are formed at distances about 20-100 event-horizon-radii \cite{jets}.
At such distances one practically is not able to distinguish the exterior field
of KBH from the exterior field of rotating objects of complete different kinds
(see for example \cite{Fiziev0902.1277} and the references therein).

As a result, the only working way to get information about
the real nature of the central engine is to study jet's spectra.
The different objects yield different frequencies of perturbations of space-time geometry,
because of the different boundary conditions on their surface (See the articles by Fiziev in \cite{QNM}
and \cite{Fiziev0902.1277}).
Thus, measuring the real jet's frequencies one can get indisputable evidences about the actual nature of the central engine.

In the present article we are probing only the KBH model for generation of jets.
The results for other possible models will be discussed elsewhere.
Due to the well known "no hair" theorems,  the non-charged KBH solutions of Einstein equations,
which potentially may be of astrophysical interest,
depend on extremely small number of parameters -- the mass and the angular momentum.
The unique boundary conditions on the event horizon yield unique robust spectra,
defined only by the mass and the angular momentum of the KBH. These spectra differ essentially from the spectra
of relativistic jets, generated by central engine of any other nature.
Thus the KBH-based model of relativistic jets seems to be easily recognizable
from observational point of view, at least up to the disguise effects by the KBH environment.

Using the Teukolsky Master Equation (TME)
we acquire the linearized perturbation of the Kerr metric (see \cite{Teukolsky}).
From there, we follow the procedure established by Teukolsky for separation
of the the variables in the TME using the substitution: $\Phi=e^{(\omega t+m\phi)i}S(\theta)R(r)$
where $m=0,\pm 1,\dots$ and $\omega=\omega_R+i\omega_I$ is a complex frequency (with $\omega_I>0$).

The exact solution of radial equation for the radial function $R(r)$:
\begin{align}
&\Delta\,{\frac {d^{2}}{d{r}^{2}}}R \left( r \right) +2\, \left( s+1\right)\left( r-
M \right){\frac {d}{dr}}R \left( r \right)  + \left( {\frac {{{\it K}}^{2}-2\,is \left( r-M \right) {
\it K}}{\Delta}}-4\,is\omega\,r-{\it \lambda} \right) R \left( r
\right)=0
\label{TRE}
\end{align}
can be expressed in terms of confluent Heun functions \cite{talks, Fiziev0902.1277, PFDS, RBPF}).

The most important for GRB physics seem to be electromagnetic waves with $s=-1$.
In this case two independent exact solutions in outer domain are:
\begin{align}
R_1 \left( r \right) ={\it C_1}\,{{ e}^{-i\omega\,r}} \left( r-{\it
r_{_+}} \right) ^{{i\frac { \omega\,({a}^{2} + {{\it r_{_+}}}^{
2})+am }{-{\it r_{_+}}+{\it r_{_-}}}}} \left( r-{\it r_{_-}} \right)
^{{ i\frac {\omega\,({a}^{2}+{{\it r_{_-}}}^{2})+am}{- {\it
r_{_+}}+{\it r_{_-}}}}+1} {\it HeunC} \left(
\alpha,\beta,\gamma,\delta ,\eta,-{\frac {r-{\it r_{_+}}}{{\it
r_{_+}}-{\it r_{_-}}}} \right) \label{R1}
\end{align}
and
\begin{align}
R_2 \left( r \right) ={ \it C_2}\,{{ e}^{-i\omega\,r}} \left( r-{\it
r_{_+}} \right) ^{{-i\frac {\omega\,({a }^{2}+{{\it
r_{_+}}}^{2})+am}{-{\it r_{_+}}+{\it r_{_-} }}+1}} \left( r-{\it
r_{_-}} \right) ^{{ i\frac{\omega\,({a}^{2}+{{\it
r_{_-}}}^{2})+am}{- {\it r_{_+}}+{\it r_{_-}}}+1}} {\it HeunC}
\left( \alpha, -\beta,\gamma,\delta ,\eta,{ -\frac {r-{\it
r_{_+}}}{{\it r_{_+}}-{\it r_{_-}}}} \right), \label{R2}
\end{align}
where: $\alpha =2\,i \left( {\it r_{_+}}-{\it r_{_-}}
 \right) \omega$,  $\beta =-{\frac {2\,i(\omega\,({a}^{2}+{{\it
r_{_+}}}^{2})+am)}{{\it r_{_+}}-{\it r_{_-}}}}-1$,
$\gamma =-{\frac {2\,i(\omega\,({a}^{2}+{{\it
r_{_-}}}^{2})+am)}{{\it r_{_+}}-{\it r_{_-}}}}+1$,\\
$\delta =-2i\!\left({\it r_{_+}}-{\it r_{_-}}
\right)\!\omega\!\left(1-i
 \left( {\it r_{_-}}+{\it r_{_+}} \right) \omega \right)$,\\
$\eta =\!\frac{1}{2}\frac{1}{{ \left({\it r_{_+}}-{\it r_{_-}}
\right) ^{2}}}\Big[ 4{\omega}^{2}{{\it r_{_+}}}^{4}+ \left(4i \omega
-8{\omega}^{2}{\it r_{_-}}\right) {{\it r_{_+}}}^{3}+ \left(
1-4a\omega\,m-2{\omega}^{2}{
a}^{2}-2A \right)  \left( {{\it r_{_+}}}^{2}+{{\it r_{_-}}}^{2} \right) + \\
 \left(4\,i\omega\,{\it r_{_-}} -8i\omega\,{\it r_{_+}}+4A-4{\omega}^
{2}{a}^{2}-2 \right) {\it r_{_-}}\,{\it r_{_+}}-4{a}^{2} \left( m+\omega\,a
 \right) ^{2} \Big]$

For the angular function $S(\theta)$, as an exact solution the angular equation:
\begin{equation}
 \left[\left(1-u^2\right)S_{lm,u}\right]_{,u}+\left[(a\omega u)^2+2a\omega su+s+{}_sA_{lm}-\frac{(m+su)^2}{1-u^2}\right]S_{lm}=0
\label{TAE}
\end{equation}
we use polynomial solutions in terms of Heun polynomials (\cite{talks, Fiziev0902.1277, PFDS, RBPF}). 
In equation \eqref{TAE} we are using the variable $u=\cos\theta$.
Putting $\Omega=a\omega$ we have explicit expressions:
\begin{align}
S_{{\pm},s,m}^{(-1)}(\theta)={{\rm e}^{\pm \Omega\,\cos\theta }}& \left(\cos\left(\theta/2\right)\right)^{\mid s-m \mid}
  \left(  \sin \left(\theta/2\right)\right) ^{- \mid s+m \mid}\times\\&
  {\it HeunC} \left( \pm 4\,\Omega,\mid\! s\!-\!m\! \mid,\mid\! s\!+\!m\! \mid,-4\,\Omega\,s,\frac{{m}^{2}-{s}^{2}}{2}+2\,\Omega\,
s-{\Omega}^{2}-A-s,  \cos^2 \frac{\theta}{2}    \right)\nonumber
\label{S1}
\end{align}
and
\begin{align}
S_{{\pm},s,m}^{(1)}(\theta)={{\rm e}^{\pm \Omega\,\cos \left( \theta \right) }}& \left(\cos \left( \theta/2\right) \right)^{\mid s-m \mid} \left(\sin \left( \theta/2 \right)\right)
 ^{- \mid s+m \mid}\times\\
&{\it HeunC}  \left( \mp 4\,\Omega,\mid\! s\!+\!m \!\mid,\mid\! s\!-\!m\! \mid,-4\,\Omega\,s,\frac{{m}^{2}-{s}^{2}}{2}-2\,\Omega\,s
-{\Omega}^{2}-A-s,  \sin^2 \frac{\theta}{2}
\right).
\nonumber
\label{S2}
\end{align}

The polynomial solutions of TME describe one way waves
and are most suitable for modeling of the relativistic jets  \cite{Fiziev0902.1277}.
Posing the polynomial condition on the angular solutions,
we obtain for the one way electromagnetic waves on Kerr background the explicit formula
\begin{equation}
A_{s=-1, m}(\omega)\!=\!-{\Omega}^{2}-2\,\Omega\,m\!\pm\!2\,\sqrt {{\Omega}^{2}\!+\!\Omega\,m}.
\label{A}
\end{equation}
This simple form of the separation constant $A$ for polynomial solutions is the most significant mathematical advantage
of our model of jets, generated by electromagnetic perturbations of Kerr metric.

The KBH boundary condition for the radial equation ca be obtained using the following assumptions:

1. On the horizon we allow only incoming in the horizon waves.
Then we obtain the different solution working in each interval of frequencies:
a) for $m=0$, only $R_2$;  b) for $m>0$ $R_1$, if $\omega_R \in (-\omega_{cr}, 0) $ and $R_2$ on the outside;
for $m<0$ $R_1$ when $\omega_R \in (0, \omega_{cr}) $ and $R_2$ outside.

2. On infinity we allow only outgoing waves.
In general, the function $R$ is a linear combination of an ingoing ($R_{\leftarrow}$)
and an outgoing ($R_{\rightarrow}$) wave:
$R=C_{\leftarrow}\,R_{\leftarrow}+C_{\rightarrow}\,R_{\rightarrow}$
with some constants $C_{\leftarrow}$ and $C_{\rightarrow}$.
In order to have only outgoing waves, we need to have $C_{\leftarrow}=0$.
This equation defines the spectral condition for the frequency $\omega$.
The main mathematical problem is that the explicit form of the constant $C_{\leftarrow}$ is not known.
Therefore to solve the spectral problem we use the around way, proposed by Fiziev in \cite{QNM}:
The straightforward check shows  that for solutions (\ref{R1}), (\ref{R2})
we have $\lim\limits_{r\to\infty}\frac{R_{\rightarrow}}{R_{\leftarrow}}= 0$
in the special direction $arg(r)=3\pi/2 - arg(\omega)$ in complex plane $\mathbb{C}_r$.
Taking $\lim\limits_{r\to\infty}\frac{R}{R_{\leftarrow}}$ in this direction,
we obviously obtain the spectral condition for $\omega$ in the form
\begin{equation}
C_{\leftarrow}=\lim\limits_{r\to\infty}\frac{R}{R_{\leftarrow}}=0.
\label{Rbc}
\end{equation}

\section{Numerical Results}

The numerical evaluations with confluent Heun functions are generally too complicated for a number of reasons.
At present the only software package able to deal with them is Maple.
Unfortunately, the existing versions of Maple package require too much time for numerical evaluation,
even on modern fast computers with large amount of memory.
Additional problem is that the procedure calculating those functions is not working well in the whole
complex plane and special attention must be paid to the regions where the procedure becomes unstable.

\subsection{Visual Resemblance of the Solutions and Real Jets}

We already reported that plotting the solutions of the angular equations with relation (\ref{A}),
we obtain figures that resemble jets (see \cite{talks,PFDS}).
Illustrations of the jet features of the model can be seen from the animations
of such oscillating solutions on our site \cite{gas}.

An interesting comparison with Nature can be obtained looking at the
picture of the discovered by NASA's Spitzer Space Telescope
``tornado-like`` object Herbig-Haro 49/50, created from the shock
waves of powerful protostellar jet hitting the circum-stellar
medium. "More observations should help us to unravel its mysterious
nature" \cite{HH49/50}. Without any doubts in this case the jet is
not related with BH. This one, as well as many other real
observations prove the presence of BH to be not necessary for
generation of jets. The collimated jets of GRB are another example
of a possible application of our simple model. There the main
problem still remains the physical nature of the central engine,
too. Clearly, if possible, a common model of jets of different
scales is most desirable.

\begin{figure}[!htb]
\centering
\includegraphics[width=300pt, height=100pt ,bb=0 0 578 259]{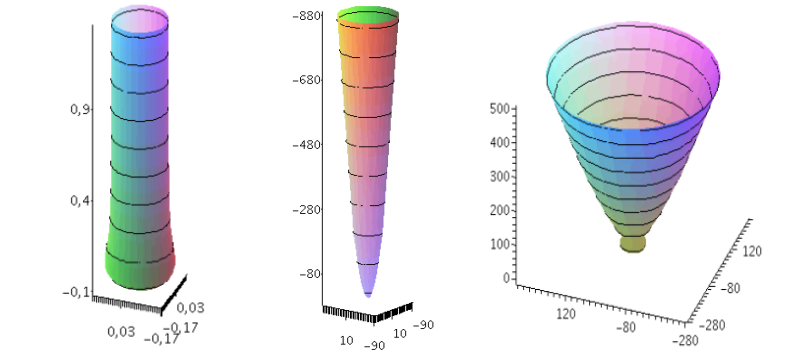}
\caption{Some of the shapes of jets that our model can represent.
Compare it with the picture of Herbig-Haro 49/50 (see
\cite{HH49/50})} \label{NASA}
\end{figure}


\subsection{Complex Super-Radiance-like Jet Modes}

The study of the solutions of radial Teukolsky equation (\ref{TRE})
is much more complicated from computational point of view.
Even after fixing the relation (\ref{A}) between A and $\omega$,
solving the KBH boundary condition (\ref{Rbc}) for
radial equation is not a straight-forward process,
because of the analytical and numerical problems connected with the evaluation of the confluent Heun function.
Our method consisted in plotting and examining the spectral condition in the complex plane $\mathbb{C}_\omega$
to find points that resemble roots of the transcendental equation (\ref{Rbc}).
Then we tested those points with more precise root-finding algorithms.

Thus we found for the case $s=-1$, $2M=1$, $a/M=0.99$, $|m|=1$ two roots:
$$\omega_{1,2}=\pm0.8676087+i\,0.1236275.$$

For the case $m=0$ we obtained another pair of roots
$$\omega_{1,2}=\pm 0.006394+i\,0.1354325801.$$
To confirm those roots, we plotted the studied function in small regions around the zeroes
and obtained the expected conus-like form of the modulus of the function in the complex domain
and the complete rotation of it's phases in angle $2\pi$ around it's simple zeroes.

Interesting in these roots is that the absolute value of their real part is precisely
the known critical frequency $\omega_{cr}=am/2Mr_{_+},m>0$
that is connected with the superradiance phenomena (for a recent review see the last reference in \cite{QNM}).
To the best of our knowledge this is the first time that an imaginary part of this quantity
is discovered solving BH boundary conditions in pure vacuum (i.e. without any mirrors, etc).
In the case of jet's solution we have a critical phenomenon around the complex value of the frequency.
It's explicit manifestation is the change of the role of solutions (\ref{R1}) and (\ref{R2}) outside
the even horizon of KBH when $|\omega_R|$ crosses the critical value $am/2Mr_{_+}$.
The complexity of the critical frequency shows an essential difference between our jets-from-KBH
solutions and standard QNM, obtained using regular solutions of the angular equation (\ref{TAE}) \cite{Teukolsky}.
It is also important that the real and the imaginary parts of the critical frequency are with the same magnitude.

The lack of zeros with $\omega_I<0$ in our numerical studies can be both due to the stability of the model,
or to numerical problems in that area.
The study of the zeroes is a field of intensive research by the team and more conclusive
results will be presented somewhere else.

\section{Conclusion}
Our simple toy model of central engine seems to be able to produce some of the basic features observed in GRB.
The Teukolsky Master Equation with the correct boundary conditions is fundamental enough to account
for all types of GRB.
The polynomial angular solutions of TME show a jet-like structure.
In the radial equation, the essential assumptions we used is that the imaginary part of the frequency
with BH boundary conditions (only entering waves on the horizon and only going to infinity waves)
should be positive. This provides stability of the solutions in direction of time-future infinity
and indicates an explosion in direction of time-past infinity.
The roots we found numerically agree with our assumption.
Interesting new result is the complex critical frequency $\omega_c^{jet}=\omega_R+i\omega_I$ ($\omega_I\sim\omega_R$)
of "superradiance" that appeared, showing  that jet's to-be-superradiance
modes decrease exponentially in time in the direction of the future and blow up in direction of the past.

\section{Acknowledgements}

This article was supported by the Foundation "Theoretical and
Computational Physics and Astrophysics" and by the Bulgarian National Scientific Found
under contracts DO-1-872, DO-1-895 and DO-02-136.

One of us (DS) is grateful to the organizers of the Fourth Aegean Summer School, 17-22 September 2007,
Lesvos, Greece for financial support of the participation in this School.


\begin{thebibliography}{}
%
\bibitem{Piran} Piran T., 1999 Physics Reports {\bf 314} 575-667
%
\bibitem{Meszaros} Meszaros P., 2001 Science {\bf 291} 79-84
%
\bibitem{Zhang} Zhang B., Meszaros P., 2002 ApJ {\bf566} 712-722
%
\bibitem{Burrows} Burrows D. N. et al., 2005 Science, {\bf 309}1833-1835
%
\bibitem{Evans} Evans et al. 2007 A\&A {\bf 469} 379
%
\bibitem{cannonball}  Dado S., Dar A., 2009 {\em The cannonball model of long GRBs - overview} astro-ph/0901.4260
%
\bibitem{Kerr} Kerr R. P., 1963 PR, {\bf 11} 237
%
\bibitem{Mirabal} Mirabal N. et al., 2006  ApJ {\bf 643} L99-L102

%
\bibitem{Andromeda} Abott  B. et all, {\em Implications for the oriign of GRB070201 from LIGO observations},  astro-ph/0711.1163
%
\bibitem{short_long} Ghirlanda G., Nava L., Ghisellini1  G., Celotti A.,  Firmani C. {\em Short versus Long Gamma–Ray Bursts: spectra, energetics,
                        and luminosities}, astro-ph/0902.0983

                        \v{R}ipa J., M\'{e}sz\'{a}ros A., Wigger C., Huja D, Hudec R., Hajdas W., 2009 {\em Search for Gamma-Ray Burst Classes
                        with the RHESSI Satellite} astro-ph/0902.1644
%
\bibitem{Teukolsky} Teukolsly S. A., 1972 PRL {\bf16}, 1114

                    Teukolsly S. A., 1973 ApJ {\bf 185}, 635-473

                    Press W. H., Teukolsly S. A., 1973 ApJ {\bf 185}, 649-673

                    Teukolsly S. A., Press W. H., 1974 ApJ {\bf 193}, 443
%
\bibitem{superradiance} Wheeler J., 1970 in {Study Week on Nucley of Galaxies}, ed D. J. K. O'Connell, North Holland, 1971
                        (Pontificae Academie Scripta Varia, {\bf 35})

                        Zel'divich Ya. B., 1971 Sov. Phys. JETP Lett {\bf 14} 180

                        Zel'divich Ya. B., 1972 Sov. Phys. JETP {\bf 62} 2076

                        Press W. H., Teukolsky S. A., 1972 Nature {\bf 238} 211

                        Starobinskiy A. A., 1973 Sov. Phys. JETP {\bf 64} 49

                        Starobinskiy A. A., Churilov S. M., 1973 Sov. Phys. JETP {\bf 65} 3

                        Wald R. M., 1974 ApJ {\bf 191} 231-233

                        Cardoso V., Dias O. J. C., Lemos J. P. S. , Yoshida S., 2004  Phys. Rev. D {\bf 70},  044039

                        Kodama H., 2008 Prog. Theor. Phys. Suppl. {\bf 172} 11-20
%
\bibitem{QNM}       Chandrasekhar S., and Detweiler S. L.,  1975 Proc. Roy. Soc.  London A{\bf 344} 441

                    Leaver E. W., 1985 Proc. Roy. Soc. London A{\bf 402} 285

                    Andersson N., 1992  Proc. Roy. Soc. London A{\bf 439} 47

                    Ferrari V., 1995 in {\em Proc. of 7-th Marcel Grossmann Meeting},
                    ed Ruffini R and Kaiser M Singapoore World
                    Scientific

                    Ferrari V., 1998 in {\em Black Holes and Relativistic
                    Stars} ed R. Wald  (Chicago: Univ. Chicago Press)

                    Kokkotas K. K., and Schmidt B. G., 1999 Living Rev. Relativity {\bf 2} 2

                    Nollert H.-P., 1999 Class. Quant. Grav.  {\bf 16} R159

                    Berti E., Black hole quasinormal modes: hints of quantum
                    gravity? {\em Preprint} gr-qc/0411025

                    Berti E., Cardoso V., Yoshida S., 2004 Phys.Rev. D {\bf 69} 124018

                   Fiziev P.~P., 2006 Class. Quant. Grav. {\bf 23} 2447-2468

                   Fiziev P.~P., 2007 Jour. Phys. Conf. Ser. {\bf 66} 012016

                   Ferrari V., Gualtieri L., 2008 Class. Quant. Grav. {\bf 40} 945-970
%
\bibitem{Blandford-Znajek} Blandford R., Znajek R., 1977 MNRAS {\bf 179} 433

                           Punsly B., Coronity F., 1990 ApJ {\bf 354} 583

                           Blandford R. D., {\em Black Holes and Relativistic Jets}, astro-ph/0110394

                           Komissarov S. S., 2008 {\em Blandford-Znajek mechanism versus Penrose process} astro-ph/0804.1912
%
\bibitem{jets}  Granot J., 2006 {\em Structure \& Dynamics of GRB Jets}, Talk given at the Conference "Challenges in Relativistic Jets"
                            Cracow, Poland, June 27, 2006, http://www.oa.uj.edu.pl/2006jets/talks.html

                            Vlahakis N., 2006 {\em Jet Driving in GRB Sources}, Talk given at the Conference "Challenges in Relativistic Jets"
                            Cracow, Poland, June 27, 2006, http://www.oa.uj.edu.pl/2006jets/talks.html

                            Marshall H.L., Lopez L.A., Canizares C. R., Schulz N. S., Kane J. F., 2006 {\em Relativistic Jets from the
                            Black Hole in SS 433}, Talk given at the Conference "Challenges in Relativistic Jets"
                            Cracow, Poland, June 27, 2006, http://www.oa.uj.edu.pl/2006jets/talks.html

                            K{\:o}enigl A., 2006 {\em Jet Launching - General Review} Talk given at the Conference "Challenges in Relativistic Jets"
                            Cracow, Poland, June 27, 2006, http://www.oa.uj.edu.pl/2006jets/talks.html
%
\bibitem{talks} Fiziev P. P., 2007  {\em Exact Solutions of Regge-Wheeler and Teukolsky  Equations},
               talk given on 23 May 2007 at the seminar of the Astrophysical Group of the Uniwersytet Jagiellonski,
               Institut, Fizyki, Cracow, Poland, http://tcpa.uni-sofia.bg/research/

               Fiziev P. P., 2007 {\em Exact Solutions of Teukolsky  Equations},
               talk given at the Conference Gravity, Astrophysics and Strings at Black Sea, 10-16 June 2007, Primorsko, Bulgaria,
               http://tcpa.uni-sofia.bg/conf/GAS/files/Plamen\_Fiziev.pdf


               Fiziev P. P., Staicova D. R., 2007 {\em A new model of the Central Engine of GRB and the Cosmic Jets},
               talk given at the Conference Gravity, Astrophysics and Strings at Black Sea 10-16 June 2007, Primorsko, Bulgaria,
               http://tcpa.uni-sofia.bg/conf/GAS/files/GRB\_Central\_Engine.pdf

               Fiziev P. P., Staicova D. R., 2007 {A new model of the Central Engine of GRB},
               talk given at the Fourth Aegean Summer School, 17-22 September 2007, Lesvos, Greece
               http://tcpa.uni-sofia.bg/research/DStaicova\_Lesvos.pdf

               Fiziev P. P., 2007  {\em Exact Solutions of Regge-Wheeler and Teukolsky  Equations},
               talk given on 28 December 2007 at the seminar of the Department of Physics, University of in Nis, Serbia,
               http://tcpa.uni-sofia.bg/research/
%
\bibitem{Fiziev0902.1277} Fiziev P. P., 2009 {Classes of Exact Solutions to Regge-Wheeler and Teukolsky  Equations}, gr-qc/0902.1277
%
\bibitem{PFDS} Fiziev P. P., Staicova D. R., 2009 {\em A new model of the Central Engine of GRB and the Cosmic Jets} astro-ph.HE/0902.2408
%
\bibitem{RBPF} Borissov R S, Fiziev P. P., 2009 {\em Exact Solutions of Teukolsky Master Equation with Continuous Spectrum} gr-qc/0902.3617
%
\bibitem{gas} GAS@BS official site: http://tcpa.uni-sofia.bg/conf/research
%
\bibitem{HH49/50} http://gallery.spitzer.caltech.edu/Imagegallery/image.php?image\_name=sig06-002

                  http://www.cfa.harvard.edu/news/2006/pr200606.html
%
\end{thebibliography}
\end{document}